\documentclass[10pt,twocolumn]{elsart}
\usepackage{graphicx}
\oddsidemargin
-10mm \evensidemargin -10mm

\newcommand{\z}{\mbox{\boldmath $\zeta$}}

\newcommand{\ks}{\mbox{\boldmath $\xi$}}

\newcommand{\rr}{\mbox{\boldmath $r$}}

\newcommand{\q}{\mbox{\boldmath $q$}}

\newcommand{\kk}{\mbox{\boldmath $k$}}

\begin{document}
\begin{frontmatter}
\title{Neutron reflection from condensed matter, the Goos-H\"anchen
effect and coherence.}
\author{V.K.Ignatovich\thanksref{mail}}
\address{Frank Laboratory of Neutron Physics of Joint Institute for
Nuclear Research FLNP JINR, 141980, Dubna Moscow region, Russia}
\thanks[mail]{ignatovi@jinr.ru}
\begin{abstract}

The Goos-H\"anchen (G-H) effect for neutron reflection from
condensed matter is considered. An experiment to quantify the
effect is proposed. The relation of G-H shift to the neutron
coherence length is considered.
\end{abstract}

\end{frontmatter}

\section{Introduction}

\begin{figure}[t]
{\par\centering\resizebox*{8cm}{!}{\includegraphics{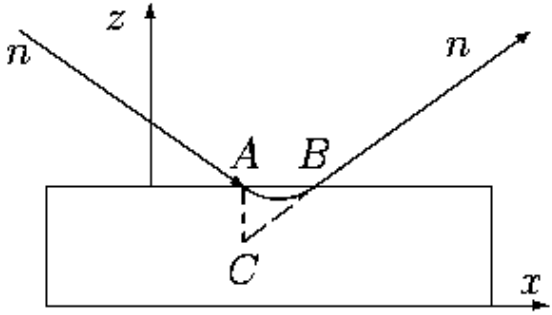}}\par}
\caption{\label{gus} Schematical explanation of the G-H shift}
\end{figure}

Newton suggested first (see, for example~\cite{lah,gu1}), that a
beam of light at a specular reflection shifts some distance along
the surface of a reflecting mirror, as shown in fig. \ref{gus}. If
the incident light is represented by a ray, then propagation of
the ray in a mirror can be described by some trajectory, where
exit point $B$ of the reflected ray does not coincide with
entrance point $A$ of the incident one. The shift $AB$ is called
G-H effect, because F. Goos and H. H\"anchen~\cite{gu1,gu2}
experimentally measured it.

The shift should exist not only for light but also for particles,
because propagation of particles in quantum mechanics is described
by wave functions similar to the wave field of the light.

The shift depends on interaction of particles with matter and
investigation of it can be useful for applied~\cite{lah,sha} and
fundamental research. In this article we consider G-H effect in
neutron physics (see, for example, \cite{maz} and references there
in). We address several problems. First, we calculate longitudinal
G-H shift at total reflection for restricted Gaussian beams (like
in optics) and for wave packets. We believe that this approach is
more general than the one (with the help of two plane waves) used
in~\cite{maz}. Second, we calculate deviation of reflected beam
from specular direction for Gaussian beam (it is similar
to~\cite{kuh} in x-ray optics) and for wave-packets. And third, we
discuss the problem of enhanced G-H effect at reflection from a
thin layer evaporated on a totally reflecting substrate. We
believe that investigation of G-H effect in optics of neutrons can
help to understand deeper the nature of the neutron wave function:
whether it is a plane wave or a wave packet. If it is the wave
packet then whether it is related to preparation of the neutron
beam or it is an intrinsic property of the neutron.

Section 2 reminds the reader how the G-H effect is calculated. We
consider plane waves, finite beams and wave packets reflection
from a single interface and from a layer on a substrate. In
section 3 we discuss how peculiarities related to the G-H effect
can be measured, and what information on neutron wave packet can
be obtained.
We address there a question whether calculation of reflection
amplitude from a layer with two interfaces, which uses coherent
superposition of coherent multiple reflection of waves (see, for
example~\cite{dar,god,par,and}), is accurate, and justify the
appropriateness of introduction in~\cite{kor} of so called
``coherence length''. This length determines such a length of the
G-H shift, at which reflection coefficient contains a considerable
contribution of incoherent reflection.

At the end of this paper, in conclusion, we shortly discuss
applicability of our calculation of the neutron G-H effect to
x-rays

\section{Calculation of the G-H shift}

Here we consider neutron reflection from an ideal surface of a
semiinfinite mirror, which means that we have a single interface.
Reflection takes place because interaction of neutron with matter.
This interaction is characterized by an optical potential $u/2$,
where $u=4\pi N_0b$, $N_0$ is atomic density, $b$ is coherent
scattering amplitude on a single atom, and for convenience we set
$\hbar=m=1$. If normal component $k_z$ of the incident neutron
momentum $\kk$ is small ($k_z^2<u$), the neutron is totally
reflected from the interface, i.e. reflection coefficient
$|R|^2=1$. If normal component $k_z$  is large ($k_z^2>u$), the
reflection is not total (in the following we call it ``nontotal'')
, and $|R|^2<1$. Here $R$ denote reflection amplitude.

Reflection coefficient depends not only on $\kk$ and $u$, but also
on structure of the wave function of the incident particle. Below
we consider three types of this function: plane wave, stationary
beam, and a wave packet.

\subsection{Plane wave}

Let the incident particle be described by a plane wave
$$\exp(i\kk_0\rr-i\omega t),$$
where $\omega=k_0^2/2$ is the energy of the neutron and
$\kk_0=(\kk_\|,-k_z)$ is a wave vector, with component $\kk_\|$
parallel, and $k_z$ --- normal to the interface.

The reflected wave is
\begin{eqnarray}
R\exp(ik_zz+i\kk_\|\rr_\|-i\omega t)=\nonumber\\
|R| \exp(-2i\phi(k_z)+ik_zz+i\kk_\|\rr_\|-i\omega t),\label{reflwave}
\end{eqnarray}
where the reflection amplitude
$$R(k_z)=\frac{k_z-k'_z}{k_z+k'_z}$$
($k'_z=\sqrt{k^2-u}$) is represented as $R=|R|\exp(-2i\phi)$.

The phase $\phi$ and the absolute value $|R|$ depend only on
$k_z$. When $k_z^2>u$ and $u/2$ is purely real, i.e. there are no
losses, then $|R|<1$ (reflection is nontotal), and $\phi=0$. Since
there is no phase shift $\phi=0$, the phase of the reflected wave
at every points on the interface is equal to the phase of the
incident wave.

When $k_z^2<u$ then $|R|=1$, i.e. reflection is total, and the
phase is nonzero: $0\le \phi=\arccos(k_z/\sqrt u)\le\pi/2$. It
means that the phase of the reflected wave (\ref{reflwave}) is
equal to the phase $k_zz+\kk_\|\rr_\|-\omega t$ of the incident
wave at points $r_\|$ displaced along the interface by a distance
$\delta r_\|=2\phi/k_\|$. This distance can be called the G-H
shift, but a spatial displacement can be well defined only for
finite size beams. In case of infinite plane waves we can talk
only about the phase shift.

\subsection{Two plane waves}

In~\cite{maz} the G-H shift was calculated with the help of two
plane waves. The incident neutron was supposed to be described by
the wave function
$$\psi_i=\exp(i\kk_1\rr)+\exp(i\kk_2\rr)$$,
consisting of two waves with wave vectors
$\kk_i=(k_{ix},k_{i\bot})$, where $k_x$ is a component
along surface, and $k_\bot$ is
perpendicular to it. This function can be also represented as
\begin{equation}
\psi_i=2\exp(i\kk\rr)\cos(\delta\kk\rr),
\label{wp3}
\end{equation}
where $\kk=(\kk_2+\kk_1)/2$,
$\delta\kk=(\kk_1-\kk_2)/2$. We see, that at the interface $z=0$
we have a plane wave $\exp(ik_xx)$ modulated by $\cos(\delta k_xx)$.

After total reflection this wave
function is transformed into
$$\psi_r=R(k_{1\bot})\exp(i\kk_1^r\rr)+R(k_{2\bot})
\exp(i\kk_2^r\rr)=$$
$$=\exp(i\kk_1^r\rr+i\phi_1)+\exp(i\kk_2^r\rr+i\phi_2),$$
where for total reflection we replaced reflection amplitudes $R(k_{i\bot})$,
which depend only on normal komponents of the wave vectors
by $\exp(i\phi_i)$. Wave vectors of the reflected waves are
$\kk_i^r=(k_{ix},-k_{i\bot})$. The sum of the reflected waves can
be also represented as
\begin{equation}
\psi_r=2\exp(i\kk^r\rr+i\phi)\cos(\delta\kk^r\rr+\delta\phi),
\label{wp4}
\end{equation}
where $\kk^r=(\kk_2^r+\kk_1^r)/2$, $\phi=(\phi_1+\phi_2)/2$,
$\delta\kk=(\kk_1^r-\kk_2^r)/2$, and $\delta\phi=(\phi_1-\phi_2)/2$.
We see, that at the interface $z=0$
we have a plane wave $\exp(ik_xx+\phi)$ modulated by
$\cos(\delta k_xx+\delta\phi)$. If we represent $\delta\phi$ as
$\delta k_x\xi$ (why not?) we obtain that modulation of the reflected
wave is shifted along $x$ axis by $\xi=\delta\phi/\delta k_x$, which
can be considered and was accepted in~\cite{maz} as the G-H shift.

\subsection{Wave function of a beam}

Imagine now that the wave function of a neutron is confined in
space around a ray, i.e. it looks like a cylindrically symmetrical
Gaussian~\cite{kuh}. Such a wave function can be stationary, and
it creates an elliptical spot at the interface~\cite{kuh}. We will
place an origin $\rr=0$ of the coordinate system in the center of
this spot and choose $xz$ to be the plane of incidence. The beam
at the interface is
$$\psi(\rr_\|,z=0,t)=\exp(ik_xx-i\omega t)\times$$
\begin{equation}\label{gus1}
\exp\left(-s^2\left[\frac{x^2\cos^2\theta}{2}+\frac{y^2}{2}
\right]\right),
\end{equation}
where $\rr_\|=(x,y)$, $\omega=k^2/2=(k_x^2+k_z^2)/2$, $1/s$ is the beam's
width, and $\cos\theta=k_z/k\equiv\sqrt{k^2-\kk^2_x}/k$ is the effective angle
of incidence.

Once we find the the wave function of the reflected beam at the
interface, the position of the center of the reflected beam with
respect to the origin will give the magnitude of the G-H shift.

To find the reflected wave function we need first to restore the
complete incident one from its spot (\ref{gus1}). In order to do
that we represent (\ref{gus1}) as a Fourier expansion
$$\psi_0(\rr_\|,z=0,t)=\int_{-\infty}^\infty \frac{d^2q_\|}{2\pi s^2\cos\theta}
\times$$
\begin{equation}\label{gus2}
\exp(i\q_\|\rr_\|-i\omega t)\exp\left(-\frac{(q_x-k_x)^2}{2s^2\cos^2\theta}-\frac{q_y^2}{2s^2}\right),
\end{equation}
and multiply the integrand by a factor $\exp(-iq_zz)$. Then,
$$\psi_0(\rr_\|,z,t)=\int_{-\infty}^\infty \frac{d^2q_\|}{2\pi s^2\cos\theta}
\e^{i\q_\|\rr_\|-iq_zz-i\omega t}\times$$
\begin{equation}\label{gus3}
\exp\left(-\frac{(q_x-k_x)^2}{2s^2\cos^2\theta}-\frac{q_y^2}
{s^2}\right),
\end{equation}
where $q_z$ must satisfy $\q_\|^2+q_z^2=2\omega=k^2$ so that the function
(\ref{gus3}) satisfies the free Schr\"odinger equation. Thus
$q_z=\sqrt{k^2-q_\|^2}$.

The incident wave function can be represented as a superposition of plane
waves, while the reflection of a plane wave is known. Thus, we can immediately
find the wave function of the reflected particle:
$$\psi_r(\rr_\|,z,t)=\int_{-\infty}^\infty \frac{d^2q_\|R(q_z)}{2\pi s^2\cos\theta}
\e^{i\q_\|\rr_\|+iq_zz-i\omega t}\times$$
\begin{equation}\label{gus4}
\exp\left(-\frac{(q_x-k_x)^2}{2s^2\cos^2\theta}-\frac{q_y^2}
{s^2}\right),
\end{equation}
where $R(q_z)$is the reflection amplitude of the incident plane wave
$\exp(i\q_\|\rr_\|-iq_zz)$. This amplitude is well known:
\begin{equation}\label{gus5}
R(q_z)=\frac{q_z-q'_z}{q_z+q'_z}=\exp(-2\chi),
\end{equation}
where $\chi=$arcch$(q_z/\sqrt u)=$arcsh$(q'_z/\sqrt u)$, and
$q'_z=\sqrt{q_z^2-u}$.

Since $q_z$ is not an independent variable, $q_z=\sqrt{k^2-q_z^2}$, the
amplitude $R$ through $q_z$ depends on $\q_\|$: $R(q_z)=\exp(-2\chi(q))$, where
$q=|\q_\||$, and
\begin{equation}
\chi(q)={\rm arcsh}(\sqrt{k^2-q^2-u}/\sqrt u).
\label{wp6}
\end{equation}

We assume for simplicity that $s$ is small, i.e. the radius $1/s$
of the beam is large. In that case the integral (\ref{gus4}) can
be easily calculated asymptotically. The function $2\chi(q)$ in
(\ref{gus4}) is expanded around maximum point $\q=\kk_\|=(k_x,0)$
of the Gaussian up to the linear term
$2\chi(q)=2\chi_0-(\q-\kk_\|)\ks$, where $\chi_0=$arcsh$(k'z/\sqrt
u)$, $\ks=(\xi_x,0)$, and
\begin{equation}
\xi_x=-2\frac{d}{dq_x}\chi(q)|_{\q=\kk_\|}=2\frac{k_x}{k_zk'_z}.
\label{wp5}
\end{equation}
When $k_z^2<u$ (total reflection) the wave-vector component $k'_z$ and thus
$\xi_x$ become imaginary, $k'_z=ik''_z= i\sqrt{u-k_z^2}$ and
$\xi_x=-i\overline{\xi}_x=-2ik_x/k_zk''_z$.

\subsubsection{Nontotal reflection}

At $k^2_z>u$ the cross-section of the reflected beam at the interface $z=0$,
according to (\ref{gus4}), is
$$\psi_r(\rr_\|,z=0)=\int_{-\infty}^\infty\frac{R(k_z)d^2q_\|}{2\pi s^2
\cos\theta}\times$$
$$\exp(i\q_\|\rr_\|+(q_x-k_x)\xi_x)\times$$
\begin{equation}\label{1gus4}
\exp\left(-\frac{(q_x-k_x)^2}{2s^2\cos^2\theta}-\frac{q_y^2}
{s^2}\right),
\end{equation}
where for simplicity we omitted the time factor $\exp(-i\omega
t)$. After integrating over $d^2q_\|$ we obtain
$$\psi_r(\rr_\|,z=0)=R(k_z)\exp\left(\frac{s^2}{2}\xi_x^2\cos^2\theta\right)\times$$
$$\exp\left(-\frac{s^2}{2}[x^2\cos^2\theta+y^2]\right)\times$$
\begin{equation}\label{gus6}
\exp(i(k_x+\xi s^2\cos^2\theta)x).
\end{equation}
One can see that there is no G-H shift, since the reflected beam
is centered at the same point as the incident beam. However the
reflection amplitude is a little bit larger than for plane waves
by the factor $\exp\left(\frac{s^2}{2}\xi_x^2\cos^2\theta\right)$,
and one can find that it is not specular. The wave vector
component $k_x$ of the reflected beam is larger than that of the
incident beam by the amount
\begin{equation}
\delta k_x=\xi s^2\cos^2\theta= 2\frac{k_xk_zs^2}{k^2k'_z},
\label{gus6a}
\end{equation}
and because of the energy conservation, the component $k_z$ must
be smaller than that of the incident wave. Thus the reflected beam
slightly turns from specular direction toward the interface.

Deviation of the reflected beam from specular direction is easily
understandable, if we take into account that the reflection
coefficient is smaller for larger $q_z$. Thus the reflected beam
is enriched with smaller $q_z$ (and larger $q_x$ because of energy
conservation). Thus the average $\q$ of the reflected beam
deviates from the specular $\kk$.

\subsubsection{Total reflection}

At $k^2_z<u$ the section of the reflected beam by the interface
according to (\ref{gus4}) becomes
$$\psi_r(\rr_\|,z=0)=\int_{-\infty}^\infty\frac{R(k_z)d^2q_\|}{2\pi s^2\cos\theta}\times$$
$$\exp\left(-\frac{(q_x-k_x)^2}{2s^2\cos^2\theta}-\frac{q_y^2}
{s^2}\right)\times$$
\begin{equation}\label{2gus4}
\exp(i\q_\|\rr_\|-i(q_x-k_x)\overline{\xi}_x),
\end{equation}
where
\begin{equation}\label{3gus4}
\overline{\xi}_x=2\frac{k_x}{k_zk''_z},\qquad
k''_z=\sqrt{u-k_z^2}.
\end{equation}
Integrating over $d^2q_\|$ we obtain
$$\psi_r(\rr_\|,z=0)=R(k_z)\exp(ik_xx)\times$$
\begin{equation}\label{1gus6}
\exp\left(-\frac{s^2}{2}[(x-\overline{\xi}_x)^2\cos^2\theta+y^2]\right).
\end{equation}
We see that the spot center of the reflected beam is shifted with respect to
that of the incident one by a distance $\overline{\xi}_x$, which is the G-H
shift. The reflection in this case is completely specular. However, the phase
$k_xx$ is identical to that of the incident wave, i.e. it does not contain the
correction $k_x\overline{\xi}_x$, which we would expect because of the G-H
shift.

Above, we limited ourselves to the linear term in the expansion of
the exponent $\chi$ in (\ref{gus5}). If we retain quadratic terms
we can find the broadening of the reflected beam. This, however,
does not bring any new insight because of the Gaussian beam own
broadening.

Expansion of $\chi(q)$ is valid only when $|k'_z|\gg s$. It is no longer valid
near $k'_z=0$ because $\xi_x$ diverges at this point. If we want to find the
reflected beam for the critical point $k'_z=0$, we have to approximate exponent
$\chi$ by a function
\begin{equation}
\chi(q)\approx \sqrt{\frac{2k_x(q_x-k_x)}{u}},
\label{wp9}
\end{equation}
and calculate the integral (\ref{gus4}) using the steepest descent
method. As a result the spot of the reflected beam at the
interface is deformed, and the reflection is not exactly specular.
This case will be analyzed elsewhere, when it will be clear hot to
study properties of the single neutron wave function
experimentally.

\subsection{The G-H shift for a wave packet}

In the previous paragraph we consider the G-H shift for a particle represented
by a stationary beam-like wave function. It is logical, however, to describe a
particle by a moving wave packet. The wave packet is not a stationary wave
function, and its Fourier representation is
\begin{equation}
\psi_0(\rr,t)=\int\limits_{}^{}A(\q)\exp(i\q\rr-i\omega_q t)d^3q,
\label{gau}
\end{equation}
where all components of the vector $\q=(\q_\|,q_z)$ are
independent variables, and $\omega_q=q^2/2$. For coefficients
$A(\q)$ we use Gaussian function,
\begin{equation}
A(\q)=\frac{1}{(2\pi s^2)^{3/2}}\exp(-(\q-\kk)^2/2s^2),
\label{gau0}
\end{equation}
where for incident particle $\kk=(\kk_\|,q_z)$

Integrating (\ref{gau}) over $\q$ gives the Gaussian wave packet
in space
$$\psi_0(\rr,t)=\frac{1}{(1+is^2t)^{3/2}}\times$$
\begin{equation}
\exp\left(-\frac{s^2}{2(1+is^2t)}(\rr-\kk_0
t)^2+i\kk_0\rr-i\omega_kt\right),
\label{gau1}
\end{equation}
where $s^2t$ characterize spreading of the wave packet. We suppose that $s$ is
small and neglect spreading.

Cross-section of the wave packet at the interface depends on time.
We can choose the time when the wave packet center crosses the
interface, $t=0$. The cross-section at that time is
\begin{equation}
\psi(\rr_\|,z=0,t=0)=
\exp\left(-\frac{s^2}{2}\rr_\|^2+i\kk_\|\rr_\|\right).
\label{gau2}
\end{equation}

The reflected wave function is
\begin{equation}
\psi_r(\rr,t)=\int\limits_{}^{}R(q_z)A(\q)\exp(i\q_r\rr-i\omega_q t)d^3q,
\label{gau3}
\end{equation}
where $\q_r=(\q_\|,q_z)$ ($q_z$ is independent variable) and $R(q_z)$ is given
by (\ref{gus5}).

\subsubsection{Total reflection}

First we consider the case of the total reflection $k_z^2<u$, when
$\chi$ in (\ref{gus5}) is $\phi=\arcsin(\sqrt{1-q_z^2/u})$. We
assume that $s$ is small and expand $2\phi(q_z)$ near $q_z=k_z$ up
to the linear term,
$2\phi(q_z)=2\phi(k_z)-\tilde{\zeta}_z(q_z-k_z)$, where
\begin{equation}
\tilde{\zeta}_z=2/k''_z=2/\sqrt{u-k_z^2}.
\label{gus6b}
\end{equation}
Placing $R(q_z)=R(k_z)\exp(i(q_z-k_z)\tilde\zeta_z)$ into (\ref{gau3}) and
taking into account (\ref{gau0}), we obtain
\begin{eqnarray}
\psi_r(\rr,t)\approx R(k_z)\int\limits_{}^{} A(\q)\times\nonumber\\
\exp(i\q_r\rr+i\tilde{\zeta}_z(q_z-k_z)-i\omega_q t)d^3q.\label{wp11}
\end{eqnarray}
Integrating over $\q$ (neglecting spreading) gives
$$\psi_r(\rr,t)=R(k_z)\times$$
\begin{equation}
\exp\left(-\frac{s^2}{2}(\rr+\tilde{\z}-\kk t)^2+i\kk\rr-i\omega_kt\right),
\label{wp7}
\end{equation}
where $\kk=(\kk_\|,k_z)$, and the vector $\tilde{\z}$ has components
$\tilde{\z}=(0,0,\tilde{\zeta}_z)$.

At the moment $t=0$, when the incident wave packet crosses the
interface, the center of the reflected wave packet is at the point
$z=-\tilde{\zeta}_z$ under the surface (depth $C$ in fig.
\ref{gus}). Thus, the real shift is not in $x$-direction. It is in
$z$-direction. However because of motion of the reflected wave
packet its center crosses the interface. It happens at the moment
$t=\tilde{\zeta_z}/k_z$, and at this moment the center of the wave
packet is located at $\rr_\|=\kk_\|\tilde{\zeta}_z/k_z$. This
location is shifted with respect to center of incident packet. So
this shift well corresponds to the intuitive understanding
illustrated in fig. \ref{gus}, and can be called the G-H shift
(\ref{3gus4}).

It is important to note that the cross-section of the wave packet at the
interface has a circular form, not an elliptical form as in the case of a beam.
Since the linear expansion is valid for $k''_z\gg s$, the G-H shift is
$\tilde{\zeta}_z\ll2/s$, and $\overline{\xi}_\|=2k_\|\tilde{\zeta}_z/k_z\ll
2k_\|/sk_z$ respectively.

\subsubsection{Nontotal reflection}

We consider now the case $k_z^2>u$, when
$\chi=$arcsh$(\sqrt{q_z^2/u-1})$ is real. Expanding $2\chi(q_z)$
near $q_z=k_z$ up to the linear term gives
$2\chi(q_z)=2\chi(k_z)+\zeta_z(q_z-k_z)$, where
$\zeta_z=2/k'_z=2/\sqrt{k_z^2-u}$. Then we place
$R(q_z)=R(k_z)\exp(-(q_z-k_z)\zeta_z)$ into (\ref{gau3}), take
into account (\ref{gau0}) and integrate the result over $\q$,
while neglecting spreading of the wave packet. As a result, the
wave function is
$$\psi_r(\rr,t)=R(k_z)e^{\zeta^2_zs^2/2}\times$$
\begin{equation}
\exp\left(-\frac{s^2}{2}(\rr-\kk't)^2+i\kk'\rr-i\omega_{k'}t\right),
\label{gau5}
\end{equation}
where $\kk'=(k_\|,k_z-\zeta_zs^2)$, and $\omega_{k'}=k'^2/2<k^2/2$.

We see that the center of the reflected wave packet crosses the interface at
the same moment $t=0$ and at the same point $\rr_\|=0$ as the incident wave
packet, i.e. there are no G-H shift. However, the momentum of the reflected
wave packet is less than that of the incident wave packet. The normal component
of the momentum is less by the amount
\begin{equation}
\delta k_z=-\zeta_zs^2,
\label{gus6c}
\end{equation}
which means non-specular reflection. The energy $\omega_{k'}$ of the {\it
reflected} wave is less than that of the incident wave. However, since the
total energy should conserve after the elastic interaction with the interface,
the {\it refracted} wave has larger energy than the incident wave.

The change in energy is the result of the independent reflections
of the different plane waves, and thus the result of the
deformation of the wave packet at the reflection. The higher is
the normal component of the component plane wave, the less is its
reflection amplitude. To avoid such deformation we must assume
that the wave packet is an intimate property of the particle as,
for example, is in the case of the de Broglie wave
packet~\cite{bro}.  For the de Broglie wave packet the energy
after the reflection does not change, and the decrease of the
normal component of the wave vector means that the direction of
the reflected wave packet is different from specular one.

\subsection{G-H shift for reflection from a layer on a substrate}

\begin{figure}[t]
{\par\centering\resizebox*{8cm}{!}{\includegraphics{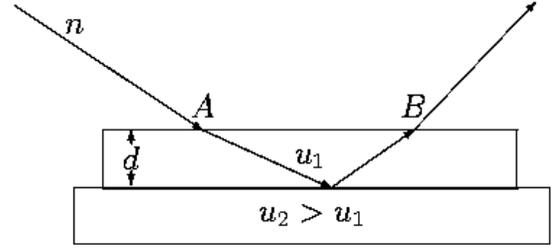}}\par}
\caption{\label{gusf} Reflection from a layer of thickness $d$
with the potential $u_1$, placed on a substrate with the potential
$u_2>u_1$. For a sufficiently thick first layer the G-H shift $AB$
can be arbitrary large.}
\end{figure}

The G-H shift for the total reflection is usually much smaller then the
dimensions of the cross-section of the wave function at the interface. Since we
assumed that $k''_z\gg s$, then
\begin{equation}
\overline{\xi}_x=2\frac{k_x}{k_zk''_z}\ll2\frac{k_x}{k_zs}\approx
\frac{2}{s\cos\theta}.
\label{wp10}
\end{equation}
The shift becomes considerably larger, if
the mirror consists of two parts as shown in fig. \ref{gusf}. It
could be, for example, a layer of Al evaporated on Be substrate.
We denote the potential of the top layer to be $u_1$ and the
bottom $u_2$, and we choose $u_1\leq u_2$.  We also choose, that
the total reflection takes place at the second interface between
the layer and substrate. We can imagine, however, that the total
reflection happens instead at the first interface of the mirror,
only this reflection happens with a large G-H shift and with deep
propagation (comparable to the thickness of the first layer) into
the mirror.

Calculations of the shift for a beam-like wave function or for a wave packet
for the mirror in fig. \ref{gusf} are the same as before. The only difference
is in the form of the reflection amplitude in (\ref{gus4}). It can be
calculated with the help of the method~\cite{and}, which takes into account
multiple reflection of plane waves from the two interfaces,
$$R(q_z)=r_{01}+\frac{(1-r_{01}^2)r_{12}\exp(2iq_{z1}d)}
{1-r_{10}r_{12}\exp(2iq_{z1}d)}=$$
\begin{equation}
\frac{r_{01}+r_{12}\exp(2iq_{z1}d)}
{1-r_{10}r_{12}\exp(2iq_{z1}d)},
\label{1gus8}
\end{equation}
where
\begin{equation}
r_{01}=-r_{10}=\frac{q_z-q_{z1}}{q_z+q_{z1}},\quad
r_{12}=\frac{q_{z1}-q_{z2}}{q_{z1}+q_{z2}},
\label{wp8}
\end{equation}
$$q_{z1,2}=\sqrt{q_z^2-u_{1,2}}.$$
We consider the case when $q_z^2-u_1\gg s^2$ and at the same time $u_2-q_z^2\gg
s^2$. In this case the reflection amplitudes from the substrate and from the
whole mirror are unit complex numbers (the losses are neglected). If we denote
$r_{12}=\exp(-2i\phi_{12})$, where
$\phi_{12}=\arcsin(q''_{z2}/\sqrt{u_2-u_1})$, $q''_{z2}=\sqrt{u_2-q_z^2}$, then
$$R(q_z)=$$
$$\e^{2iq_{z1}d-2i\phi_{12}}\frac{1+r_{01}\exp(-2iq_{z1}d+2i\phi_{12})}
{1+r_{01}\exp(2iq_{z1}d-2i\phi_{12})}=$$
\begin{equation}
\exp(2iq_{z1}d-2i\phi_{12}-2i\phi_{02})
\label{1agus8}
\end{equation}
where $\phi_{02}=$
\begin{equation}
\arcsin\left(\frac{r_{01}\sin(2q_{z1}d-2\phi_{12})}
{\sqrt{1+r^2_{01}+2r_{01}\cos(2q_{z1}d-2\phi_{12})}}\right).
\label{gau6}
\end{equation}
In the case of a beam-like wave function, expanding the phase
$2q_{z1}d-2\phi_{12}-2\phi_{02}$ in (\ref{1agus8}) over $\q_\|-\kk_\|$, where
$\kk_\|=(k_x,0)$, up to the linear term gives the G-H shift,
$$\overline{\xi}_x=\left[\frac{2k_xd}{k_{z1}}+\frac{2k_x}{k_{z1}k'_{z2}}
\right]\times$$
\begin{equation}
\frac{1+r_{01}\cos(2q_{z1}d-2\phi_{12})}
{1+2r_{01}\cos(2q_{z1}d-2\phi_{12})+r^2_{01}}.
\label{gau7a}
\end{equation}
The first term of the expression in the brackets is $k_zd$ times
larger than the shift at a single interface. The factor outside
the brackets depends on the neutron wave length and varies in the
range from $1/(1+r_{01})$ to $1/(1-r_{01})$.

Same result can be obtained in the case of a wave-packet wave function.

\section{Possible experiments to measure the neutron G-H effect}

In this section we estimate the value of the G-H effect and look into
perspectives of its measurement.

\subsection{Nonspecular reflection from a single interface}

In case of a nontotal reflection of a beam, the component
$k_x=k\sin\theta$ ($\theta$ is the angle of incidence) of the wave
vector in the reflection plane along the interface increases by
$2k_xk_zs^2/k^2k'_z$(\ref{gus6a}). This is equivalent to rotating
of the wave vector by a small angle $\gamma$:
$$k'_x\equiv k_x+\delta k_x=k_x+2k_xs^2/k_zk'_z=$$
\begin{equation}
k\sin(\theta+\gamma)=k_x\cos\gamma+k_z\sin\gamma.
\label{gau7}
\end{equation}
From (\ref{gau7}) follows that $\delta k_x=k_z\gamma$, and when comparing the
latter with (\ref{gus6a}) we find
\begin{equation}\label{4gus4}
\gamma=2\frac{k_xs^2}{k^2k'_z}.
\end{equation}
In order to estimate $\gamma$ we need to find $s$.

The width $s$ of the wave packet was estimated in~\cite{uig}. It
was found that the anomalously high loss coefficient of ultracold
neutrons in storage vessels can be explained, if the wave function
of a neutron is represented by the de Broglie wave packet with the
width $s\approx4\cdot10^{-5}k$, where $k$ is the neutron wave
number. We use this estimate in the rest of this article.

With the above estimate for $s$, if we take $k_z, k_x\approx\sqrt
u=4\cdot10^2$ (which is typical for thermal neutrons) the angle is
$\gamma\approx1.3\cdot10^{-6}$. This value is too small to
measure. However, it is possible to design an experiment with
$k'_z\approx 0.1\sqrt u$, then $\gamma\approx 13\cdot10^{-6}$ rad,
which can be measured.

In the case of the wave packets the change of the wave vector is determined by
(\ref{gus6c}), $\delta k_z=2s^2/k'_z$. If this change is the result of the
rotation by the angle $\gamma$, then $\gamma=2s^2/k_xk'_z$. Thus, using the
same estimates as in the previous paragraph we obtain the same number for the
angle $\gamma$.

\begin{figure}[t]
{\par\centering\resizebox*{8cm}{!}{\includegraphics{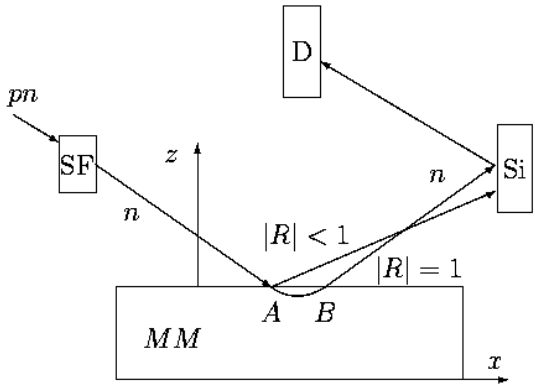}}\par}
\caption{\label{gusf1} Scheme of experiment to check deviation of
reflected neutrons from specular direction when reflection is
partial. Monochromatic polarized neutrons (pn) go through spin
flipper (SF) and are reflected from a magnetized mirror (MM). If
SF is switched off then neutron polarization is parallel to the
mirror magnetization, reflection is total and specular. The
neutrons reflected from MM go to Si single crystal and after Bragg
reflection to detector D. When SF is switched on, neutron
polarization becomes opposite to magnetization of MM and the
reflection is not total, thus the Bragg condition is not satisfied
for neutrons going to Si crystal. In order to restore Bragg
condition the Si crystal must be turned by an angle $\gamma$. The
purpose of the experiment is to measure this angle.}
\end{figure}

We can check this result with the help of an experiment, schematic
of which is shown in fig. \ref{gusf1}. In the proposed experiment,
the reflection of the monochromatic polarized neutrons from a
magnetic mirror at a fixed angle is measured. After reflecting
from the magnetic mirror, the polarized neutrons go to a single
crystal where the Bragg condition is satisfied. After reflecting
from the crystal, the neutrons are registered by a detector. When
neutrons are polarized parallel to the magnetization of the
mirror, they are totally and specularly reflected from it. If the
polarization is opposite, the reflection is partial and not
specular. The reflected beam from the mirror does not completely
satisfy Bragg condition at the single crystal, so in order to
restore the Bragg condition we need to rotate the crystal. Thus
the experiment consists of measuring rocking curves of a single
crystal for two neutron polarizations. The centers of the two
rocking curves will be shifted by $\gamma$. Of course, to achieve
sufficient precision we need incident beam also monochromatized by
an identical single crystal. More over, to avoid some false
effects the analysis, may require reflection from two specially
arranged single crystals. This can be discussed for a concrete
reflectometer.

\subsection{G-H shift at total reflection}

Let us estimate the G-H shift (\ref{3gus4}) at the total reflection from a
single interface. For thermal neutrons the ratio $k_x/k_z$ at the total
reflection is about 400. Thus the G-H shift is 400 times larger than
penetration depth $1/k''_z$, which can be estimated to be $1/\sqrt u\approx
10^{-6}$ cm. Therefore, for thermal neutrons the G-H shift is in the order of 8
$\mu$m.

For thermal neutrons with $\lambda=2$ \AA\ the spatial width of
the wave packet is $<1\mu$m, which is of the order of magnitude
{\it less} than the G-H shift. However this changes, if the
incident neutron instead of the wave packet is described by a beam
of the same width. In this case we must compare the G-H shift to
$1/s\cos\theta$, which is one order of magnitude {\it more} than
the G-H shift.

In the case of the reflection from a layer of thickness, say 1
$\mu$m, on a substrate the G-H shift is in the order of 0.8 mm and
is considerably larger even than $1/s\cos\theta$.

\begin{figure}[t]
{\par\centering\resizebox*{8cm}{!}{\includegraphics{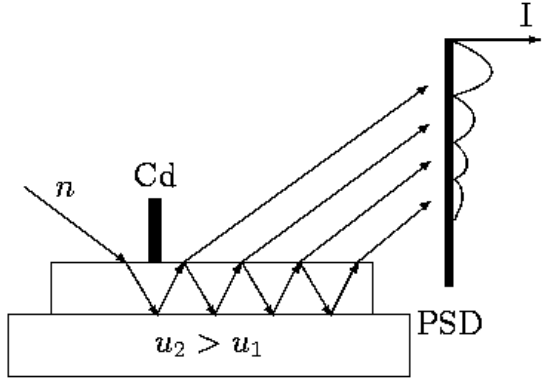}}\par}
\caption{\label{gusf2} Multiple reflection from two interfaces of
sufficiently thick film, when shifted wave fields on the mirror
surface do not overlap, can be represented as multiple incoherent
reflection. In that case a diffraction pattern can be observed at
the position sensitive detector PSD. Every maximum will correspond
beam, which escaped the film after consecutive reflections from
the substrate.}
\end{figure}

This shift is possible to measure in the experiment shown in fig.
\ref{gusf2}. If a Cadmium (Cd) plate restricts the area of the
mirror enlightened by the incident beam, then the detector will
detect only those particles, which can dive under the Cd plate.

It is possible that the beam which passed under the Cd plate will
experience multiple reflections from the vacuum-layer interface
and from the layer-substrate interface. In this case the
distribution of the neutrons at the position sensitive detector
(PSD in fig. \ref{gusf2}) will look like a diffraction pattern. In
reality, however, this pattern will be smeared by the
non-monochromaticity and by scattering from the interface
roughness and from the inhomogeneities of the film. However the
contribution of scattering can be estimated and separated, if we
can measure intensity of neutrons scattered outside of the
incidence plane.

\subsection{G-H effect and coherence}

Reflection from two or more interfaces presents a problem, which
is worth while to carefully investigate. In such a reflection we
find coherence, incoherence and a transition between them.

Indeed, expression (\ref{1agus8}) is the result of a coherent
superposition of plane waves instantaneously reflected multiple
(infinite) times from two interfaces. This expression is obtained
as a sum of geometric series
$$R=r_{01}+
t_{10}e^{iq_{z1}d}r_{12}e^{iq_{z1}d}t_{01}+$$
\begin{equation}\label{1cgus8}
t_{10}\sum\limits_{n=1}^{\infty}\left[e^{iq_{z1}d}r_{12}e^{iq_{z1}d}
r_{10}\right]^ne^{iq_{z1}d}r_{12}e^{iq_{z1}d}t_{01},
\end{equation}
where every term represents a possible path of the wave between
the moment it enters the mirror and the moment it exits the
mirror. The wave entering the mirror is defined by the
vacuum-to-layer transmission amplitude $t_{01}=2q_z/(q_z+q'_z)$,
and the wave exiting the mirror is defined by layer-vacuum
transmission amplitude $t_{10}=2q'_z/(q_z+q'_z)$. If we sum these
multiple reflections, and use the resulting reflection amplitude
(\ref{1agus8}) to find reflected wave packet, we obtain a single
exit point for the packet. If we treat every term of the sum
separately and use series (\ref{1cgus8}) to calculate reflection
of the wave packet, we obtain many reflected wave packets each
with different exit point. So the single G-H shift transforms into
multiple shifts. The coordinate of the $n$-th exit point is
proportional to $nd$. So the question is, what the multiple
reflections are in reality: are they coherent sum of plane waves
or a sequence of incoherent reflections? If they are the coherent
sum, which can be expected, because of observable interference,
then what do multiple exit points mean?

It is interesting to check whether different reflected wave
packets interfere with each other, or the interference is present
only when the spots of the consecutively reflected waves on the
interface overlap (and is absent when they do not overlap). If
they do not interfere when the spots do not overlap, then multiple
reflections become incoherent. If we can measure the reflection
with continuously changing overlapping, we can see how coherence
transforms into incoherence.

To check the phenomenon mentioned in the previous paragraph, the
experiment depicted in fig. \ref{gusf2} can be used. If the
reflected beam is originated at one point close to the one of
incidence, then it will be blocked by a cadmium shutter, and no
neutrons will be detected by the PSD. It will prove the coherent
summation of the reflected amplitudes. However, if the multiple
reflections proceed according to (\ref{1cgus8}), the distribution
of the neutrons on the PSD will be similar to a diffraction
pattern.

There is also a different way to check the transition of the coherence into the
incoherence. Suppose the reflection from the substrate is not total, i.e.
$r_{12}$ is a positive real number less than unity. In that case the full
reflection amplitude (\ref{1agus8}) is not a unit complex number
$$R(q_z)= \frac{r_{01}+r_{12}\exp(2iq_{z1}d)}
{1+r_{01}r_{12}\exp(2iq_{z1}d)}=$$
\begin{equation}
|R(q_z)|\exp(i\varphi(q_z)).
\label{1bgus8}
\end{equation}
Here
\begin{equation}
|R|=\sqrt{\frac{r^2_{01}+2r_{01}r_{12}\cos(2q_{z1}d)+r^2_{12}}{1+
2r_{01}r_{12}\cos(2q_{z1}d)+r^2_{01}r^2_{12}}}<1,
\label{gau10}
\end{equation}
and $\varphi(q_z)=$
$$\arcsin\left(\frac{r_{12}\sin(2q_{z1}d)}
{\sqrt{r^2_{01}+2r_{01}r_{12}\cos(2q_{z1}d)+r^2_{12}}}\right)-$$
\begin{equation}
\arcsin\left(\frac{r_{01}r_{12}\sin(2q_{z1}d)}
{\sqrt{1+2r_{01}r_{12}\cos(2q_{z1}d)+r^2_{01}r^2_{12}}}\right).
\label{gau11}
\end{equation}
If we expand $\varphi(q_z)$ and $\ln(|R(q_z)|)$ around the point
$\q_\|=\kk_\|$, then the G-H shift $\overline{\xi}_x$ is a complex number. Real
part of it defines the shift, and imaginary part defines the deviation of the
reflected beam from the specular direction.

\begin{figure}[t]
{\par\centering\resizebox*{8cm}{!}{\includegraphics{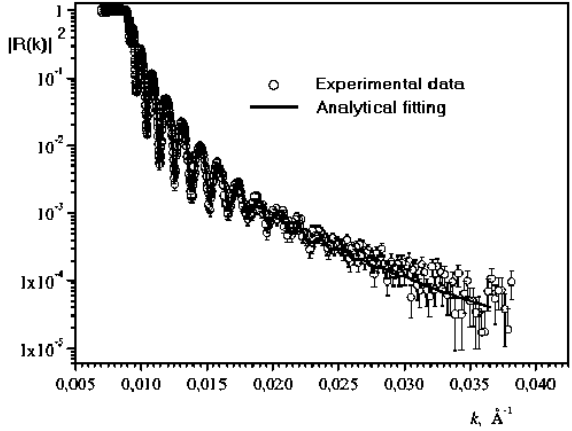}}\par}

\caption{\label{gusb}Experimentally measured reflection curve for a film on a
substrate~\protect\cite{kor}, and its theoretical fitting}
\end{figure}

A reflectometer can be used to measure the reflection amplitude dependence on
energy of the incident neutrons for a given incidence angle. Example of such
dependence is shown in fig. \ref{gusb} (reprinted from ~\cite{kor}) .
The most interesting feature of this curve is the contrast of
oscillations. Period of oscillations is related only to the width
of the film, but the contrast depends on many factors, such as
smearing of interfaces because of the surfaces
roughness~\cite{kor2}, losses because of absorption, scattering on
roughnesses and inhomogeneities, and possible contribution of
incoherent reflection.

In the experiment~\cite{kor} the contribution of the described factors was
found by fitting the theoretical curve (solid line in fig. \ref{gusb}) to
experimental points. The smearing (for theory see, for example~\cite{kor2}) was
described by two parameters related to two interfaces. The contribution of the
incoherent reflection was described by the coherence length parameter. The
smearing parameter characterizes the height of roughness. The coherence length
characterizes a certain separation of enter and exit spots of the neutron wave
function, at which the coherent and incoherent contributions to the reflection
coefficient become equal. The fitting in the experiment~\cite{kor} was done
with the coherence length and without it, and it was found that $\chi^2$ of the
fitting improves considerably, if the coherence length is included.

We show now how the coherence length, denoted $\xi_0$, was included in
theoretical description of the reflection curve shown in fig. \ref{gusb}.

Let us introduce a function $f(x)$, which monotonously increases from zero at
$x=0$ to one at $x=\infty$, and represent the reflection coefficient as the sum
\begin{equation}
|R|^2=(1-f(x))|R_c|^2+f(x)R_i,
\label{gau12}
\end{equation}
where $R_c$ is coherent reflection amplitude, and $R_i$ is the incoherent
reflection coefficient. If $x=\overline{\xi}_x/\xi_0$, then for small $x$, i.e.
for a small separation $\overline{\xi}_x$ compared to the coherence length
$\xi_0$, the reflection is completely coherent. For a large separation
$\overline{\xi}_x$ compared to the coherence length $\xi_0$, the reflection is
completely incoherent. In~\cite{kor} the error function
$\Phi(\tilde{\xi}_x/\sqrt2\xi_0)$ is chosen to be $f(x)$. Thus the reflection
coefficient is defined as
$$|R|^2(k_\bot)=
|R_c|^2(k_\bot)\left(1-\Phi\left(\frac{\tilde{\xi}_x(k_\bot)}
{\sqrt2\xi_0}\right)\right)+$$
$$R_i(k_\bot)\Phi\left(\frac{\tilde{\xi}_x(k_\bot)}
{\sqrt2\xi_0}\right),$$ where $k_\bot$ is the component of the incident wave
vector normal to the mirror, $R_c$ is given by (\ref{1agus8}), and $R_i$ is the
incoherent reflection coefficient
$$R_i=\frac{R_1+R_2-2R_1R_2}{1-R_1R_2}.$$
Here $R_{1,2}$ are reflection coefficients from the two interfaces. In the case
of ideal interfaces
$$R_1=\left|\frac{k_{\bot}-k_{1\bot}}{k_{\bot}+k_{1\bot}}\right|^2,\quad
R_2=\left|\frac{k_{1\bot}-k_{2\bot}}{k_{1\bot}+k_{2\bot}}\right|^2.$$ In the
case of smeared interfaces, when smearing is described by the Eckart
potential~\cite{kor2},
$$u(x)=\frac{u}{1+\exp(x/\sigma)}$$ with smearing parameter $\sigma$,
$$R_1=\left|\frac{\sinh(\pi \sigma_1(k_{\bot}-k_{1\bot}))}
{\sinh(\pi \sigma_1(k_{\bot}+k_{1\bot}))}\right|^2,$$
$$R_2=\left|\frac{\sinh(\pi \sigma_2(k_{1\bot}-k_{2\bot}))}
{\sinh(\pi \sigma_2(k_{1\bot}+k_{2\bot}))}\right|^2,$$
where $\sigma_{1,2}$ are smearing parameters of two interfaces.

The fitting of the experimental data revealed that the coherence
parameter $\xi_0$ is equal to 1.5 mm. Further research has to be
done to understand the meaning of this coherence parameter, and
how it relates to the width of the neutron wave packet.

\section{Conclusion}

We discussed the G-H effect in case of neutrons, however almost
everything said above is applicable also to x-rays. The most
important difference is that we do not know how to estimate the
wave packet width for x-rays. The experiments similar to those
proposed in this article may give at least an upper limit of this
very interesting parameter.

\section*{Acknowledgement}

The author is grateful to R.G\"ahler for discussions, to F.Radu
for useful remarks, to F.V.Ignatovich for help, and to
E.P.Shabalin for support.

\end{document}